\documentclass[prb,twocolumn,superscriptaddress,longbibliography]{revtex4-2}
\usepackage{amsmath,amssymb,bm}
\usepackage[utf8]{inputenc}
\usepackage{hyperref}
\usepackage{graphicx}
\usepackage{epstopdf}
\usepackage{latexsym}
\usepackage{subfigure}
\usepackage[usenames, dvipsnames]{color}
\usepackage[usenames, dvipsnames]{xcolor}
\usepackage{natbib}
\usepackage{braket}
\usepackage{float}
\usepackage[normalem]{ulem}
\usepackage{comment}
\usepackage{mathtools}
\usepackage{array}
\usepackage{multirow}
\usepackage{chemformula}
\usepackage{svg}

\newcommand\redsout{\bgroup\markoverwith{\textcolor{red}{\rule[0.5ex]{2pt}{0.4pt}}}\ULon}
\newcommand\bluesout{\bgroup\markoverwith{\textcolor{blue}{\rule[0.5ex]{2pt}{0.4pt}}}\ULon}

\newcommand{\prlsec}[1]{{{\it #1:--}}}
\newcommand{\pibytwo}{\frac{\pi}{2}}

\newcommand{\unhighlight}[1]{{\color{black}{#1}}}

\newcommand{\SPhide}[1]{{}}

\begin{document}
\title{A solvable embedding mechanism for one-dimensional spinless and Majorana fermions in higher-dimensional spin-$\frac{1}{2}$ magnets}
\author{Sumiran Pujari}
\affiliation{Department of Physics, Indian Institute of
Technology Bombay, Mumbai, MH 400076, India}
\affiliation{Max Planck Institute for the Physics of Complex Systems, 01187 Dresden, Germany}
\email{sumiran.pujari@iitb.ac.in (on sabbatical
leave)}

\begin{abstract}
We write down a class of two-dimensional quantum spin-$\frac{1}{2}$ Hamiltonians 
whose eigenspectra are exactly solvable via the Jordan-Wigner 
transformation.  
The general structure corresponds to a suitable grid composed of 
XY or X-Ising spin chains and Z-Ising spin chains and 
is generalizable to higher dimensions.
They can host stacks of one-dimensional spinless fermion 
liquids with 
gapless excitations and power-law correlations coexisting with 
ordered spin moments (localized spinless fermions). 
Bond-dependent couplings thus can be an alternate mechanism 
than geometric frustration of $SU(2)$-symmetric couplings to obtain
spinless fermionic excitations.
Put in a different way, bond-dependent couplings allow for 
an embedding of one-dimensional
spinless fermion (Tomonoga-Luttinger) liquids and solids
and also Majorana excitations in 
higher dimensions.
They can accommodate a  simpler set of site-local conserved quantities apart
from the more intricate, interlocked set of plaquette-local or bond-local conserved quantities 
in Kitaev's honeycomb model with Majorana excitations.
The proposed grid structure may provide
an architecture for quantum engineering with controllable qubits.

\end{abstract}
\maketitle

\section{Introduction}
\label{sec:intro}
Emergent fractionalized excitations are a remarkable manifestation
of strongly correlated many-body physics.
A canonical example is the one-dimensional ($1d$) 
spin-$\frac{1}{2}$ antiferromagnetic 
chain with fermionic spinon excitations~\cite{Sachdev_book1}
which are colloquially said to be 
``half" of regular 
(bosonic) spin-wave excitations or magnons 
in presence of magnetic ordering. Magnons being close
to product quantum states with low 
quantum entanglement 
allow one to form a classical image as
quanta of smooth 
undulations of the ordered magnetic moments
like phonons for lattice vibrations
and photons for electromagnetic waves.
Spinons on the other hand may not have simple
classical images. They are generally speaking
a complicated superposition of the underlying 
spin states with high quantum entanglement.
In $1d$ one can have images of spinon excitations as 
dressed defects of some underlying order, e.g. 
$1d$ transverse field quantum Ising model 
or the valence bond solid phase of the 
$1d$ Ghosh-Majumdar model where the spinon excitations
can be imagined as a dressed domain wall excitation.
However, this image does not easily extend to other
liquid-like situations especially in $d>1$.
Such emergent
excitations are thus fascinating apart from the 
qualitative new effects they engender.

Finding solvable models where such physics is explicit
is of importance as proofs-of-principle. They
serve as counterparts to field theoretic arguments that rely
on low-energy approximations, perturbative renormalization
group treatments and universality~\cite{Sachdev_book}. The downside is that
solvability is often restricted to points
in the phase diagram~\cite{measure_zero_footnote}. 
The Kitaev honeycomb model is 
a famous solvable model of fractionalized 
(free fermionic) Majorana excitations~\cite{Kitaev_2006,Baskaran_Sen_Shankar_PRB_2008, Fu_Knolle_Perkins_PRB_2018}
on top of a spin liquid (SL) ground state
in $2d$ $S=\frac{1}{2}$ antiferromagnets.
In this work, 
we will write down a set of
exactly solvable spin-$\frac{1}{2}$ models with 
bond-dependent couplings which host spinless fermionic 
or Majorana excitations 
akin to the Kitaev model but with rather different physics.
The technological application possibilities 
due to the new physics exist both for materials engineering and
artifical quantum technologies. They are discussed 
in the final Sec.~\ref{sec:conclu} (in particular
Sec.~\ref{subsec:physical_realization}).

To motivate our line of thought, 
take an infinite XY spin chain 
\unhighlight{$H=J \sum_{\langle i,j \rangle} \left( S^x_i S^x_j + S^y_i S^y_j \right)$ as shown in 
Fig.~\ref{fig:single_off_chain} (a).
Following standard notation, $i$,$j$,$\ldots$ refer to lattice sites,
$S^\mu_i$ refer to 
spin-$\frac{1}{2}$ angular momentum operators on site $i$ and
$c^\dagger_i$,$c_i$ refer to fermionic creation and 
destruction operators at site $i$.
This is solved using the well-known Jordan-Wigner(JW) mapping~\cite{Jordan_Wigner_1928}:
\begin{align}
c^\dagger_i = \prod_{j<i} (-1)^{S^z_j+1/2} S^+_i \nonumber \\
c_i = \prod_{j<i} (-1)^{S^z_j+1/2} S^-_i \nonumber    \\
n_i \equiv c^\dagger_i c_i = S^z_i + \frac{1}{2} \nonumber
\end{align}
where
$S^\pm_i = \frac{S^x_i \pm S^y_i}{2}$ often represented
by Pauli $2\times2$ matrices.
This map respects
the mutual consistency of fermion anticommutation
algebra $c_i c^\dagger_j + c^\dagger_j c_i = \delta_{ij}$
and spin angular momentum algebra 
$S^\alpha_i S^\beta_j - S^\beta_j S^\alpha_i = i 
\delta_{ij} \epsilon_{\alpha \beta \gamma} S^\gamma_i$
where $\delta$, $\epsilon$ are the Kronecker
and Levi-Civita symbols respectively. 
The fermionic vacuum is chosen here to be
$\prod_i \otimes |\downarrow_i\rangle$.
The geometry of ``$j<i$'' in the above map
defines the JW-string.
After the mapping, the XY chain becomes the
free fermions hopping chain
$H=\frac{J}{2} \sum_{\langle i,j \rangle} c^\dagger_i c_j + \text{h.c.}$.

\begin{figure}
    \centering
    (a) \includegraphics[width=0.7\linewidth]{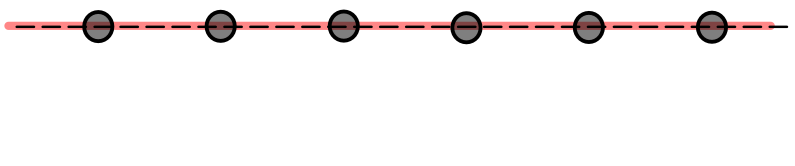}
    (b) \includegraphics[width=0.7\linewidth]{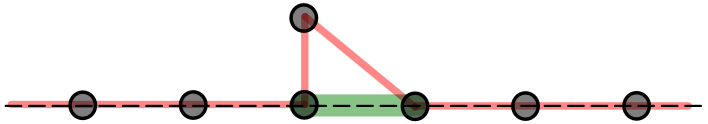}

    \vspace{4mm}
    (c) \includegraphics[width=0.7\linewidth]{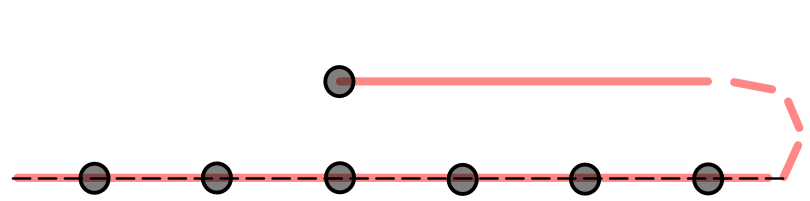}
    \caption{\unhighlight{Gray dots represent $S=\frac{1}{2}$ degrees
    of freedom, black dashed lines represent the XY couplings,
    and the light red line represents the Jordan-Wigner string respectively. 
    \unhighlight{Panel (a) shows a part of an 
    infinite XY chain. Panels (b) and (c) show 
    the same XY chain along with 
    an additional single, disconnected spin for two
    different choices of the Jordan-Wigner string 
    geometry.}
    }}
    \label{fig:single_off_chain}
\end{figure}

Now consider an additional single, disconnected 
``off-chain" spin as in Fig.~\ref{fig:single_off_chain} (b).
The full solution is simply the product state
of the disconnected spin and  
the JW-solved Fermi sea,
i.e.
$|\psi\rangle = |\sigma\rangle \otimes |\text{JWSL}_c\rangle$,
$|\text{JWSL}_c\rangle = \prod_{\epsilon_p \leq 0} c^\dagger_p
|\Omega\rangle$ and
$\epsilon_p = J \cos (p)$
where $p$ refers to reciprocal momentum.
}
Consider the situation when the off-chain
spin is included in the JW-string
as in Fig.~\ref{fig:single_off_chain} (b). 
If the
off-chain spin is in $|\downarrow\rangle$ state, 
then the XY chain
maps to spinless fermions hopping as before
with uniform hopping amplitudes on all bonds. 
If it is in $|\uparrow\rangle$ state 
(occupied state), 
then hopping amplitudes are not uniform. 
The green-highlighted
bond in Fig.~\ref{fig:single_off_chain} (b) has 
oppositely signed hopping amplitude compared to the 
rest of the bonds. 
However this change of sign in hopping amplitudes
is a pure gauge artifact in $1d$. One can remove it by the 
following transformation: 
$c_i \rightarrow d_i$ on sites left of the green bond, 
and $c_i \rightarrow -d_i$
to the right.
This is special to $1d$ and 
hopping amplitude signs 
can not be ``gauged away" in $d>1$.
They rather specify fluxes in the elementary
plaquette units in higher dimensions.
Thus for any $|\sigma\rangle = a\:|\uparrow\rangle + 
b\:|\downarrow\rangle$ for the disconnected spin, the solution is
$\left( a\:|\uparrow\rangle \otimes |\text{JWSL}_c\rangle + b\:|\downarrow\rangle \otimes
|\text{JWSL}_d\rangle \right)$
since $|\text{JWSL}_c\rangle$ and $|\text{JWSL}_d\rangle$ 
are degenerate.

Of course, one may just not include the 
disconnected spin in the JW-string,
i.e. put it at the ``very end" as shown in 
the bottom panel of Fig.~\ref{fig:single_off_chain}
to not affect the JW-hopping amplitudes,
which gives $|\sigma\rangle \otimes |\text{JWSL}_c\rangle$.
We are not bothering with formal issues
akin to the Hilbert's paradox of the Grand Hotel, 
i.e. finite size chains can be
handled without any essential difference to the argument 
as the results will show.
The above argument can be applied to any number of disconnected off-chain spins.
Thus off-chain spins can bring in only pure gauge
artifacts if they are themselves not involved kinetically 
with the spin chain. 
Once off-chain spins are involved kinetically, 
then JW-solvability generally breaks down. 

\unhighlight{
The rest of the paper is organized as follows:
In Sec.~\ref{sec:model}, the basic design
principle behind the solvable models is given.
Sec.~\ref{sec:results} describes the results
obtained for different categories of model constructions 
as outlined in Sec.~\ref{sec:model}. Sec.~\ref{sec:conclu}
summarizes the various magnetic states found in the previous
sections including discussions on their
stability when going beyond the solvable parameter
regime (Sec.~\ref{subsec:stability}), 
possibilities for physical realization 
and some imagined technological applications (Sec.~\ref{subsec:physical_realization}).
}

\begin{figure*}
    \centering
    {\Large(a)}\includegraphics[width=0.38\linewidth]{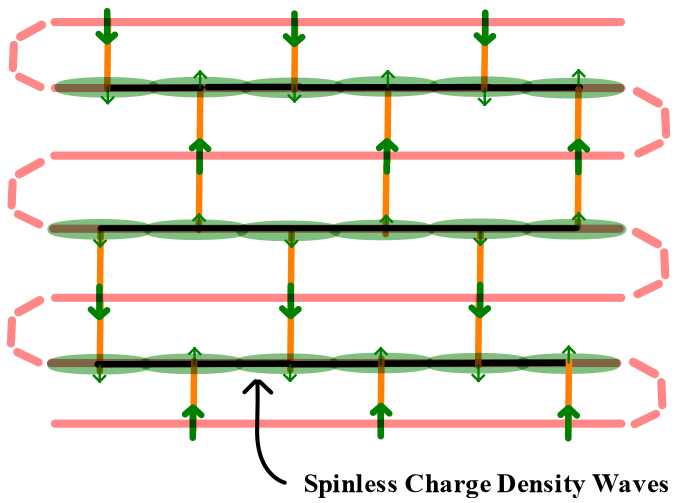} \;\;\;\;\;\;\;\;\;\;\;\;
    {\Large(b)}\includegraphics[width=0.33\linewidth]{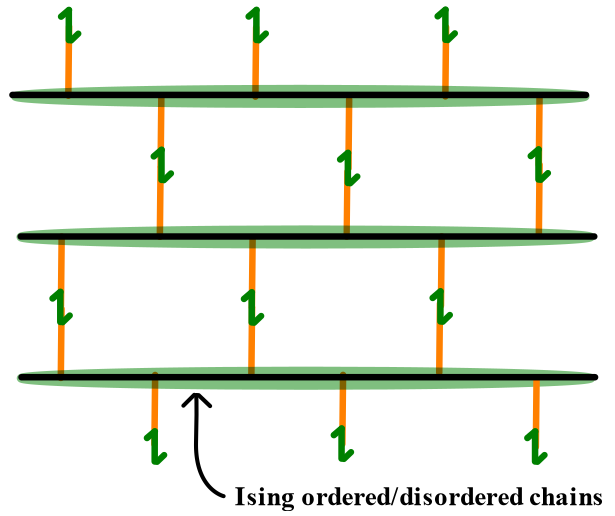}
    \caption{Example of a lattice without triangular motifs.
    Black lines represent XY and X-Ising couplings in 
    panels (a) and (b) respectively.  
    Orange lines represent Z-Ising couplings. The light red line is 
    the snake-like choice for the JW-string throughout this work.
    The ground state sector corresponds to an antiferromagnetic
     ((a); see Sec.~\ref{subsec:unfrus_xy} for further details), 
     paramagnetic ((b); see Sec.~\ref{subsec:unfrus_XX} for further details) arrangement of the conserved off-chain spins with gapped
     on-chain spinon excitations.
    The difference in the size of the off-chain
          spins and on-chain spins in the panel (a) illustrates the 
         following: since off-chain spins are conserved,
         the $Z$-moment size is maximal, while the on-chain
         $Z$-spins fluctuate due to the XY terms thus
         leading to a reduced moment size compared to the
         off-chain spins.
    }
    \label{fig:unfrustrated_xy_xx}
\end{figure*}

\section{Model construction}
\label{sec:model}

The class of models that we will write down now
avoid the ``off-chain kinetics" as
in the gedanken discussed
in the previous Sec.~\ref{sec:intro}.
The model construction will consist of collection 
of spin chains with on-chain kinetic terms 
and off-chain spins that are restricted 
to have (``interaction") Z-Ising couplings 
with the on-chain spins. 
For such a bond-dependent scenario, 
the JW-transformation can be used fruitfully and
gauge artifacts of the above sort 
are removed by a judicious ``snake-like" 
choice for the JW-string 
(Fig~\ref{fig:unfrustrated_xy_xx} (a)). 
This leads to 
a $2d$ grid of XY/X-Ising and Z-Ising 
chains
that may be written in general as 
\begin{equation}
    H = \sum_{\langle i,j \rangle_{\alpha}} 
    J_{{\langle i,j \rangle}_{\alpha}} \left( S^x_i S^x_j + \alpha 
    S^y_i S^y_j \right) 
    + \sum_{\langle i,j \rangle_{\text{Z-I}}} 
    J_{{\langle i,j \rangle}_{\text{Z-I}}} S^z_i S^z_j 
    \label{eq:gen_ham}
\end{equation}
where $\alpha=1$, $\alpha=0$ correspond to XY and X-Ising couplings
respectively. The set of $\{\langle i,j \rangle_\alpha\}$ 
and $\{\langle i,j \rangle_{\text{Z-I}}\}$ bonds
will be indicated by black and orange solid lines 
respectively in the subsequent 
diagrams. 
\unhighlight{The above models belong to the more general class
of JW-solvable models as laid out in
Refs.~\cite{Ogura_etal_2020,Chapman_Flammia_2020} 
and related papers~\cite{Minami_2016,Minami_2017,Minami_Yanagihara_2020,
Chapman_Elman_Flammia_2021,Chapman_Elman_Mann_2023,Fendley_Pozsgay_2024}.}
This embedding can also be extended to $d>2$ if the
restriction on kinetic terms to on-chains is
respected.
We will restrict ourselves to uniform couplings,
i.e. $J_{{\langle i,j \rangle}_{\alpha}} = J_{\alpha}$,
$J_{{\langle i,j \rangle}_{\text{Z-I}}} = J_{\text{Z-I}}$. 
Under the JW map we get
\begin{align}
    H = & \; \frac{J_\alpha}{2} \sum_{\langle i,j \rangle_{\alpha}} 
    \left( c^\dagger_i c_j + (1-\alpha) 
    c^\dagger_i c^\dagger_j + \text{h.c.}\right)  \\
    &    + J_{\text{Z-I}} \sum_{\langle i,j \rangle_{\text{Z-I}}} 
    \left(n_i - \frac{1}{2}\right)\left(n_j - \frac{1}{2}\right) \nonumber
    \label{eq:gen_JWham}
\end{align}
Through the above construction, $S^z_k$ or $n_k$ on an off-chain site $k$
become conserved $c$-numbers, i.e. effectively classical
(Ising or $Z_2$) variables, that converts $H$ into a quadratic
form thereby leading to the solution of the
full eigenspectrum.

Bond-dependent couplings in
spin-$\frac{1}{2}$ Hamiltonians have assumed importance since 
Jackeli and Khaliullin's seminal 
work~\cite{Jackeli_Khaliullin_PRL_2009}.
The proposed class of bond-dependent models can thus 
form a new material class 
apart from so-called Kitaev materials~\cite{Kitaev_Materials_Trebst_2017,Kitaev_Materials_Winter_etal_2017,Kitaev_Materials_Hermanns_etal_2018,Kitaev_Materials_Takagi_etal_2019,Kitaev_Materials_Trebst_Hickey_2022,Kitaev_Materials_Kim_etal_2022}. 
The models naturally divide into geometrically
bipartite lattices containing
no triangular motifs, and geometrically 
non-bipartite lattices containing
triangular motifs.
The Hamiltonians are however frustrated in all cases
due to the competition of the on-chain 
XY/X-Ising and Z-Ising terms similar
to the Kitaev model.
A further sub-division concerns if
the on-chain terms are of XY or X-Ising type.
In the following, we will consider ferromagnetic signs
for the couplings without any loss of generality. 
There are local unitaries that can give all possible
sign combinations and resultant ground states.


\section{Results}
\label{sec:results}

\unhighlight{
This section is organized as follows:
Sec.~\ref{subsec:numerics} gives a concise description
of the numerical methods used to obtain the results
in the following subsections. Sec.~\ref{subsec:unfrus_xy}
discusses the case of bipartite XY model construction,
followed by in Sec.~\ref{subsec:frus_xy_xx} the cases of non-bipartite XY and X-Ising
model constructions.
Then a subsection is devoted to analytical arguments
for perturbative stability of the magnetic ground 
states found in the previous subsections. The climax of this section
is the bipartite X-Ising model
construction in Sec.~\ref{subsec:unfrus_XX} which
hosts an intriguing liquid-like ground state.
}

\subsection{Numerical computations}
\label{subsec:numerics}

\begin{figure*}
    \centering
    {\Large(a)} \includegraphics[width=0.42\linewidth]{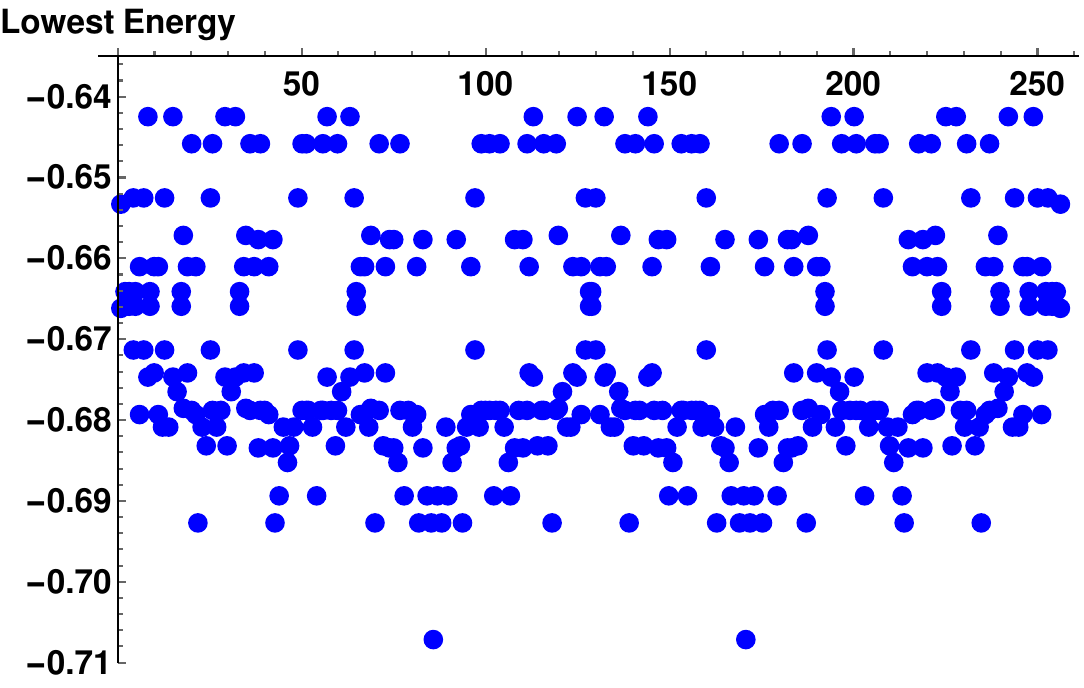} \;\;\;\;\;\;\;\;
    {\Large(b)}\includegraphics[width=0.42\linewidth]{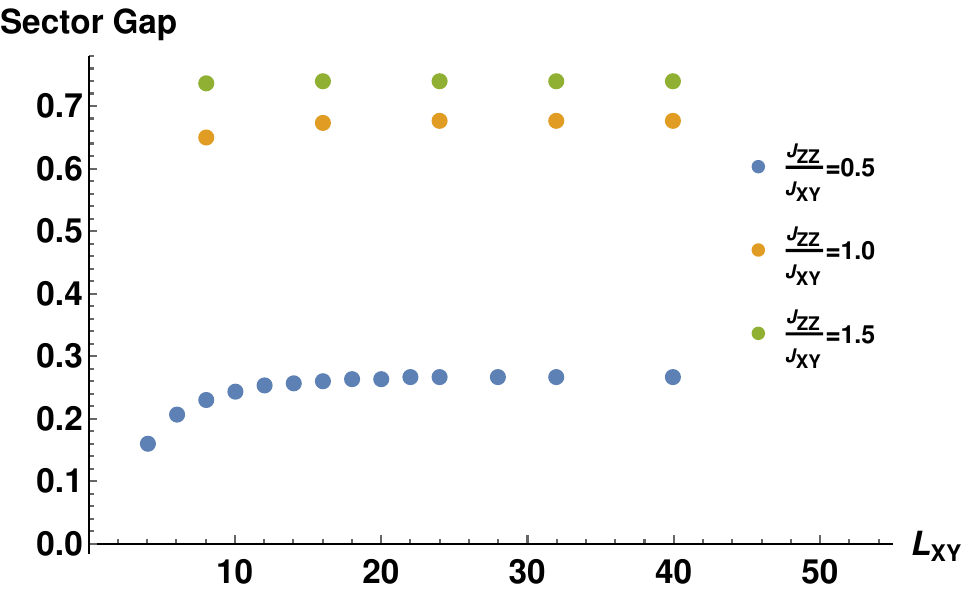}

    \vspace{0.5cm}
    {\Large(c)} \includegraphics[width=0.38\linewidth]{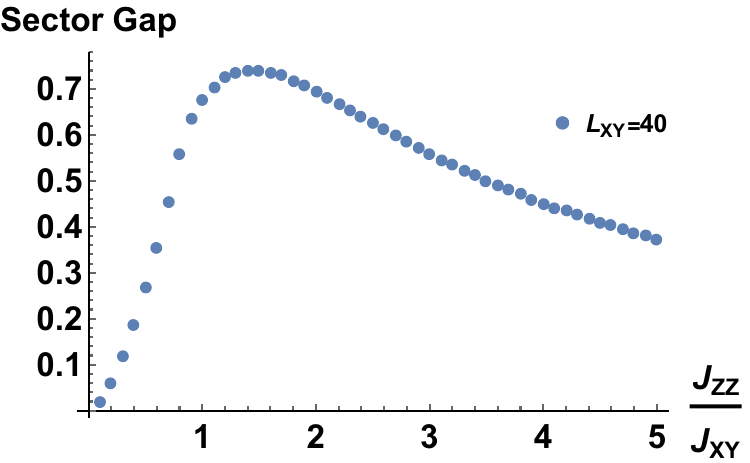}
    \caption{Panel (a) shows the lowest energies (per site 
    \unhighlight{and in units of $J_{\text{XY}}=1.$}) from each conserved off-chain
    spin sector for an $L=8+8$ chain~\ref{subsec:numerics}
    for the bipartite XY model with $\frac{J_{\text{Z-I}}}{J_{\text{XY}}}=0.5$ (Fig.~\ref{fig:unfrustrated_xy_xx} (a); 
    see Sec.~\ref{subsec:unfrus_xy} for further details). 
    The X-axis indexes
    the different sectors through the decimal equivalent of the binary-valued off-chain spin configuration. 
    One can verify that the two lowest energies corresponds to the two antiferromagnetic
    configurations for the off-chain spins
    \unhighlight{($85_{10} \equiv 01010101_2$ and $170_{10} \equiv 10101010_2$)}. Panel (b) shows the finite gap
    between the lowest energies from other sectors and the ground state energies
    for different system sizes. Panel (c) shows how this sector gap evolves
    with $\frac{J_{\text{Z-I}}}{J_{\text{XY}}}$ for $L=40$.}
    \label{fig:unfrustrated_sector_gaps}
\end{figure*}
We first briefly discuss the details of the
numerical computations that are presented in 
the following sections. 
For numerical purposes, we can focus on just
one set of on-chain spin chain and its corresponding
off-chain spins and transfer the result to the
rest of the spin chains. In other words, the exact
solution applies to infinite extent in the perpendicular
direction to the spin chains.
Mathematica codes based on them
are available at this \href{https://www.dropbox.com/scl/fo/wcirm0vmfpjap2o2wmf6p/AP8bjsdbjc5X_lB2ciF4NK4?rlkey=jgdocqipsts17xrgrs18y8u20&st=lcay0h2t&dl=0}{link}.
The codes also have further annotations that
will be of help to the interested reader.

Starting from Eq.~\ref{eq:gen_JWham} after 
JW-transforming Eq.~\ref{eq:gen_ham}, we fix the
values of the conserved off-chain spins' $S^z$
quantum numbers to either $+\frac{1}{2}$ or 
$-\frac{1}{2}$, or equivalently the conserved
occupation number of the off-chain JW-fermions'
$n$ to $0$ or $1$. For an off-chain of length $L$,
there are $2^L$ different configurations for the
conserved off-chain spins. 
Each off-chain configuration 
effectively provides a background on-site potential
to the on-chain JW-fermions, i.e.
\begin{align}
    H = & J_\alpha \sum_{\langle i,j \rangle_{\alpha}} 
    \left( c^\dagger_i c_j + (1-\alpha) 
    c^\dagger_i c^\dagger_j + \text{h.c.}\right)  \\
    &    + J_{\text{Z-I}} \sum_{\langle i,j \rangle_{\text{Z-I}}} 
    \left(n_i - \frac{1}{2}\right)\left(\langle n_j \rangle - \frac{1}{2}\right) \nonumber
    \label{eq:gen_quadham}
\end{align}
where it is understood that depending on the lattice geometry
$\langle n_j \rangle$ in the
second term above corresponds to the conserved off-chain degrees of 
freedom with value either $0$ or $1$.

For the XY cases, the above quadratic fermionic Hamiltonian 
for the on-chain spin chain of length $L$
can now be easily diagonalized in the single-particle sector for each
conserved off-chain sector. This is computationally just a $L \times L$
matrix diagonalization. The lowest energy state in each
sector is found by summing all the negative single-particle 
eigenvalues in that sector.
To find the overall many-body ground state, one needs to find that 
sector which hosts highest magnitude of the above negative
single-particle sum. For small chain lengths, this can be done
exhaustively as in the provided codes. 
For chains of bigger lengths,
the exponential growth of $2^L$ makes it difficult to an exhaustive
search. One can take advantage of translation symmetry here for
homogeneous couplings to restrict the set of conserved off-chain
configurations to those ones that are not related by translation
symmetry. This has also been shown in the provided codes.

For the X-Ising cases which lack $U(1)$ symmetry or total $\sum_i S^z_i$
conservation (equivalently $\sum_i n_i$), one needs to follow
the standard Bogoliubov-de Gennes set-up~\cite{BdG_notes_2024} 
to find the single-particle
states in this ``superconducting'' situtation with no particle
conservation but only fermion parity conservation.
Computationally, this only leads to a doubling of the 
size of the single-particle Hamilonian since now we have to
handle both the particle and hole channels together. The
additional cost of diagonalizing a $2L \times 2L$ is 
negligible and does not become a bottle-neck.
Through these computations, one can verify
the various ground states discussed in the main text apart 
from the perturbative arguments for their selection
discussed later.

\subsection{Bipartite XY} 
\label{subsec:unfrus_xy}

An example of this construction
is shown in Fig.~\ref{fig:unfrustrated_xy_xx} (a). 
The ground state is found to be an antiferromagnet in the 
Z-direction. The JW-spectrum on the chains is
gapped implying exponentially decaying transverse XY
correlations as well. The ground state can also 
be thought of as a spinless fermion charge density wave. 
In contrast to the Kitaev model where one takes 
recourse to Lieb's
theorem~\cite{Lieb_1994} to rigorously pin down the ground state sector, 
here we are not aware of any rigorous way of proving the
ground state sector selection. 
As shown in Fig.~\ref{fig:unfrustrated_sector_gaps} (a),  
one numerically finds that the ground state sector indeed corresponds
to an antiferromagnetic arrangement of the conserved off-chain
spins (sketched in Fig.~\ref{fig:unfrustrated_xy_xx} (a)). 
Such an arrangement
corresponds to on-site potentials with alternating sign, i.e.
modulating at wavevector $\pi$, for
the on-chain JW-fermions which leads to gap opening 
near the Fermi level (at $\pm \pi/2$) along with
charge density wave ordering at the same wavevector. 
Ground state sector selection can be argued for
perturbatively in 
the $\frac{J_{\text{XY}}}{J_{\text{Z-I}}} \gg 1$
and $\frac{J_{\text{XY}}}{J_{\text{Z-I}}} \ll 1$ limits. 
For $\frac{J_{\text{XY}}}{J_{\text{Z-I}}} \ll 1$,
the on-chain spin aligns with 
its partner off-chain spin to satisfy the 
ferromagnetic Z-Ising coupling. 
The neighboring on-chain spins now prefer to anti-align
to gain energy through virtual hops. A 
different argument can also be made
for this sector selection 
in the other $\frac{J_{\text{XY}}}{J_{\text{Z-I}}} \gg 1$ limit
~\cite{Subhro_private}.
These perturbative stability arguments
in the solvable regime are further discussed
in Sec.~\ref{subsec:pert_arg}.
\begin{figure*}
    \centering
    {\Large (a)} \includegraphics[width=0.38\linewidth]{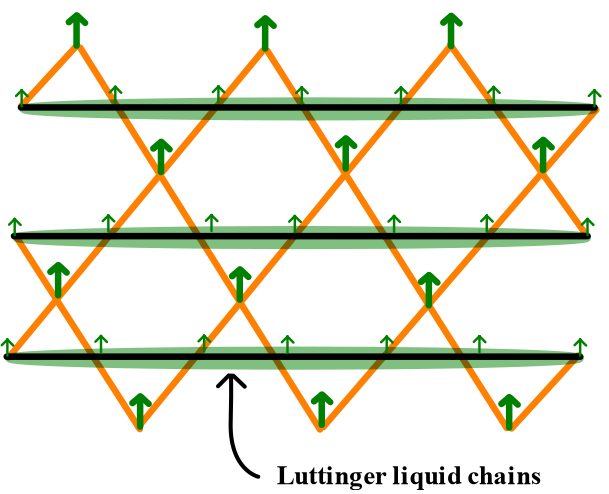} \;\;\;\;\;\;\;\;\;\;\;\;
    {\Large (b)} \includegraphics[width=0.38\linewidth]{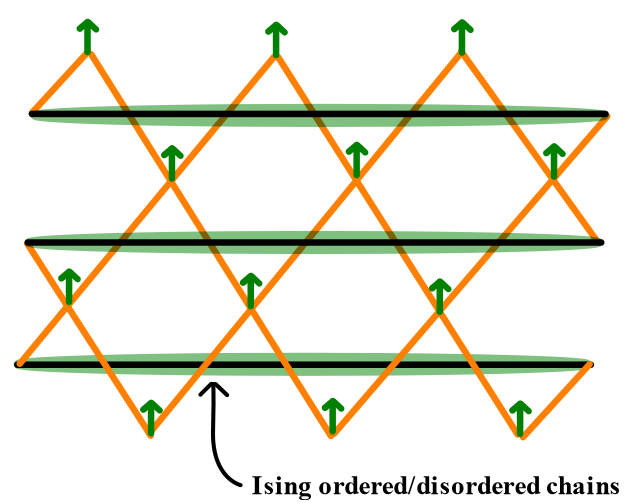}
    \caption{Example of a lattice with triangular motifs.
    Black lines represent XY and X-Ising couplings
    in panels (a) and (b) respectively.
    Orange lines represent Z-Ising couplings.
    The ground state sector corresponds to a ferromagnetic
    arrangement of the conserved off-chain spins for both cases, 
    with gapless on-chain spinon excitations in (a), and gapped
    on-chain Majorana excitations in (b). 
    See~\ref{subsec:frus_xy_xx}  for further details
    The difference in the size of the off-chain
          spins and on-chain spins again 
          illustrates the 
         difference in the $Z$-moment size 
         of the off-chain and on-chain spins
         as discussed in the caption of Fig.~\ref{fig:unfrustrated_sector_gaps}.}
    \label{fig:frustrated_xy_xx}
\end{figure*}
The ground state interpolates smoothly between the
two limits with the ratio of gap opening size to
spinless Fermi sea bandwidth being controlled
by $\frac{J_{\text{XY}}}{J_{\text{Z-I}}}$.
Upon monitoring the lowest energy states in other
conserved off-chain spin sectors, one finds a finite
energy gap as well as shown in Fig.~\ref{fig:unfrustrated_sector_gaps} (b).
It is also numerically observed that the
off-chain spin configuration corresponding to the 
lowest excited sector changes once as
we increase $\frac{J_{\text{Z-I}}}{J_{\text{XY}}}$ from domain wall 
of size one to domain wall of size two 
in the off-chain antiferromagnetic configuration
as in Fig.~\ref{fig:unfrustrated_sector_gaps} bottom. 
It is not clear if this is just energetics or 
something deeper.
Thus this construction leads to gapped antiferromagnetic
ground states.
A rigorous proof of this fact is desirable.
The X-Ising variant in Fig.~\ref{fig:unfrustrated_xy_xx} (b)
of this construction
will be addressed later in Sec.~\ref{subsec:unfrus_XX}.

\begin{figure*}
    \centering
    {\Large (a)} \includegraphics[width=0.42\linewidth]{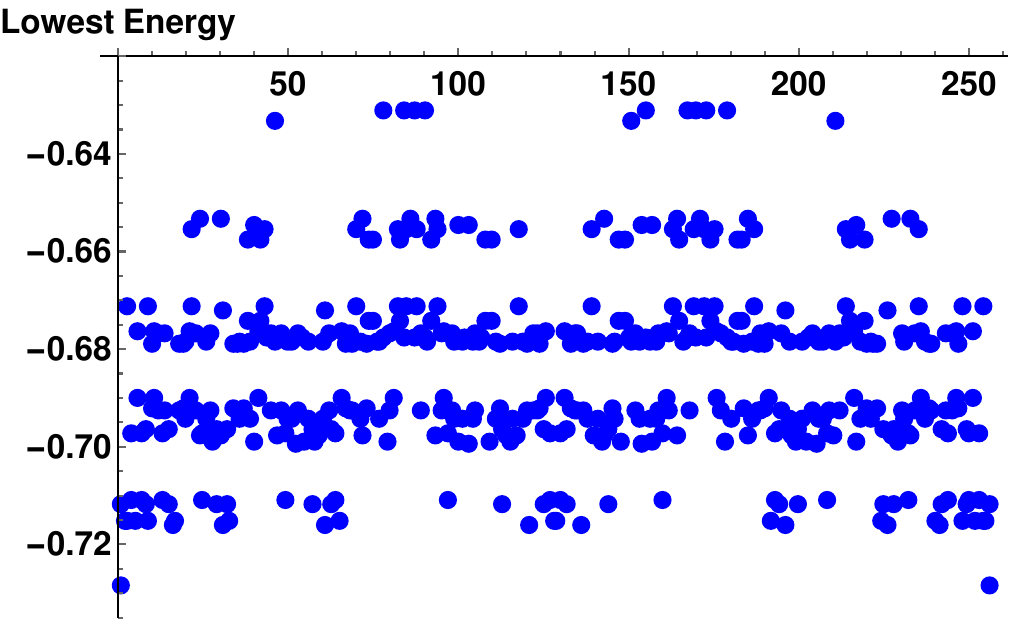} \;\;\;\;\;\;\;\;
    {\Large (b)} \includegraphics[width=0.42\linewidth]{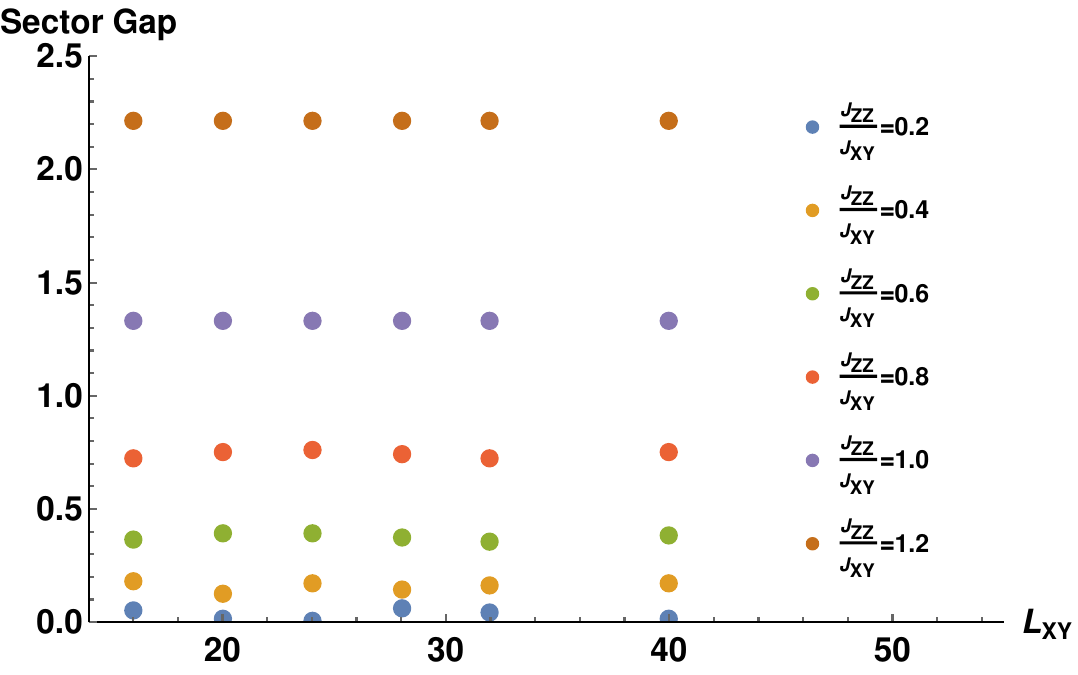}

    \vspace{0.5cm}
    {\Large (c)} \includegraphics[width=0.35\linewidth]{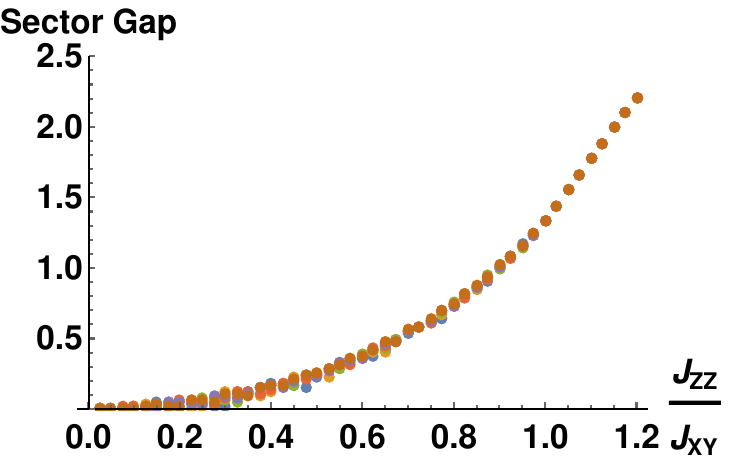}
    \caption{Panel (a) shows the lowest energies (per site \unhighlight{and in units of $J_{\text{XY}}=1.$}) from each conserved off-chain
    spin sector for an $L=8+8$ chain~\ref{subsec:numerics}
    for the non-bipartite XY construction 
    with $\frac{J_{\text{Z-I}}}{J_{\text{XY}}}=0.5$
    (Fig.~\ref{fig:frustrated_xy_xx} (a); see
    Sec.~\ref{subsec:frus_xy_xx} for further details). 
    The X-axis indexes
    the different sectors through the decimal equivalent of the binary-valued off-chain spin configuration. 
    One can verify that the two lowest energies corresponds to the two ferromagnetic
    configurations for the off-chain spins
    \unhighlight{($0_{10}=00000000_2$ and $255_{10}=11111111_2$)}. 
    Panel (b) shows the finite gap
    between the lowest energies from other sectors and the ground state energies
    for different system sizes. Panel (c) shows how this sector gap evolves
    with $\frac{J_{\text{Z-I}}}{J_{\text{XY}}}$ for these system sizes.}
    \label{fig:frustrated_sector_gaps}
\end{figure*}

\subsection{Non-bipartite XY and X-Ising}
\label{subsec:frus_xy_xx}

An example of this construction
is shown in Fig.~\ref{fig:frustrated_xy_xx} (a).
Its ground state turns out to be a ferromagnet in the 
Z-direction as evidenced in Fig.~\ref{fig:frustrated_sector_gaps} (a). 
These constructions are in fact solvable points
in the general Kagome antiferromagnet phase diagram!
The JW-fermion 
spectrum on the chains now
remains gapless implying algebraically decaying 
transverse XY correlations. \unhighlight{This 
can be thought of as a quasi-long-range-ordered
form of the so-called $U(1)$-symmetry breaking
spin superfluid order in higher dimensions in the literature~\cite{Spin_Superfluid_2008}.} 
Again the above is determined numerically.
Here too, we can make similar
perturbative arguments in $\frac{J_{\text{XY}}}{J_{\text{Z-I}}} \gg 1$
and $\frac{J_{\text{XY}}}{J_{\text{Z-I}}} \ll 1$ limits for the
ground state sector selection.
There is an interpolation 
between the fully polarized state for 
$\frac{J_{\text{Z-I}}}{J_{\text{XY}}} > 1$ and  
a half-filled spinless Fermi sea 
as $\frac{J_{\text{Z-I}}}{J_{\text{XY}}} \rightarrow 0$.
Upon monitoring the lowest energy states in other
conserved off-chain spin sectors, one again finds 
a finite energy gap as shown in 
Fig.~\ref{fig:frustrated_sector_gaps} (b).
It is also numerically observed that the
off-chain spin configuration corresponding to the 
lowest excited sector keeps changing as
we increase $\frac{J_{\text{Z-I}}}{J_{\text{XY}}}$ till
$\frac{J_{\text{Z-I}}}{J_{\text{XY}}}=1$
as in Fig.~\ref{fig:frustrated_sector_gaps} bottom.
It is not clear if this is just energetics or 
something deeper. Gaplessness and finite sizes
further complicate matters for smaller values
of $\frac{J_{\text{Z-I}}}{J_{\text{XY}}}$ in this
non-bipartite XY case. Such complications were absent in the
bipartite XY case in Sec.~\ref{subsec:unfrus_xy}.
Thus we find that this construction leads to ferromagnetic
ground states with gapless on-chain excitations. 
A rigorous proof of the ground state sector
selection is again desirable.
The X-Ising variant of this construction
has related physics (Fig.~\ref{fig:frustrated_xy_xx} (b)).
The ground state corresponds
to a ferromagnetic configuration of the off-chain
spins which effectively provides a transverse field
to the on-chain spins. Thus depending on 
$\frac{J_{\text{Z-I}}}{J_{\text{X-I}}}$, an Ising ordered
or disordered state obtains for the on-chain spins 
which in
terms of the JW fermions would be a topologically
non-trivial or trivial superconducting state
respectively with gapped Majorana excitations
and the transition at 
$\frac{J_{\text{Z-I}}}{J_{\text{X-I}}}=0.5$ on this (kagome) lattice.
The sector gaps also remain finite 
like in the non-bipartite XY variant.

\subsection{Perturbative arguments for ground state sector
selection}
\label{subsec:pert_arg}

\begin{figure}
    \centering
    \includegraphics[width=0.75\linewidth]{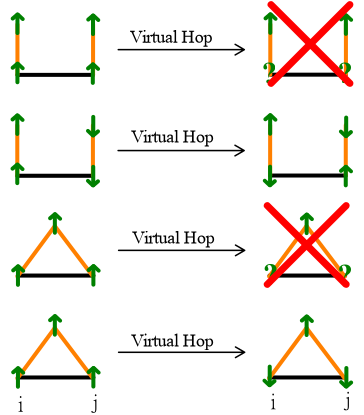}
    \caption{\unhighlight{Schematic of the 
    virtual hopping processes 
    in the $\frac{J_{\text{XY}}}{J_{\text{Z-I}}} \ll 1$ limit
    for bipartite XY, X-Ising and non-bipartite XY, X-Ising
    cases from top to bottom.
    See the beginning of Sec.~\ref{subsec:pert_arg} for the details.
    }}
    \label{fig:pert_arg_ising_limit}
\end{figure}

The $\frac{J_{\text{XY}}}{J_{\text{Z-I}}} \ll 1$ or Ising
limit for the
bipartite XY case was alluded to earlier in Sec.~\ref{subsec:unfrus_xy}. 
The top
two panels of Fig.~\ref{fig:pert_arg_ising_limit} summarize 
this argument. An analogous argument can be
made for the non-bipartite XY and X-Ising cases as shown 
in the lower panels of Fig.~\ref{fig:pert_arg_ising_limit}.
The virtual hops are caused by the XY or X-Ising 
on-chain couplings. Note for the
non-bipartite XY case, there is no virtual hopping 
energy gain to
be had due to the triangular nature of the elementary motifs as 
in the third panel of Fig.~\ref{fig:pert_arg_ising_limit}. 
For the non-bipartite X-Ising case,
virtual hopping energy can be gained from the superconducting
part of the X-Ising couplings as in the fourth panel of 
Fig.~\ref{fig:pert_arg_ising_limit} while keeping the 
on-chain short-range Ising correlations intact. 

In the opposite XY limit for $\frac{J_{\text{Z-I}}}{J_{\text{XY}}} \ll 1$,
one can argue by considering one on-chain spin chain and
its partner off-chain spins. Let us start with the bipartite XY case.
The lattice Hamiltonian is
\begin{equation}
    H = \frac{J_{\text{XY}}}{2} \sum_{\langle i,j \rangle} c^\dagger_i c_j + \text{h.c.}
        + J_{\text{Z-I}} \sum_i \left( n_i - \frac{1}{2} \right) \langle S^z_i \rangle  
\end{equation}
where $\langle S^z_i \rangle$ are the conserved off-chain partners
of on-chain JW-fermions for the lattice site $i$. The first term
equals $\sum_p \epsilon(p) c^\dagger_p c_p$ with $p \in (-\pi,\pi]$.
In the XY limit, we can focus our attention to the modes near the
Fermi points $p_F = \pm \frac{\pi}{2}$ of the unperturbed many-body
ground state which is the half-filled spinless Fermi sea, i.e.
\begin{align}
    c_i \sim \sum_{-\Lambda < \delta k < \Lambda} 
    \frac{1}{\sqrt{L}} \left( e^{i \left(\pibytwo + \delta k \right) r_i}  c_{\delta k,r} 
    + e^{i \left(-\pibytwo + \delta k \right) r_i}  c_{\delta k,l} 
    \right).
\end{align}
where $\Lambda \ll \pi$ is a cutoff scale much smaller than the lattice
scale.
$l$ and $r$ stand for left-propagating and right-propagating
modes or ``helicities" respectively, i.e. their 
corresponding group velocities
are $\propto -|\delta k|$ and $\propto +|\delta k|$ respectively.
With the above, the perturbation term can be evaluated to
\onecolumngrid
\begin{widetext}
\begin{align}
    \sum_i \langle S^z_i \rangle \left(n_i - \frac{1}{2}\right) = 
    \sum_{-\Lambda < \delta k,\delta k' < \Lambda} 
    \frac{1}{L} \Bigg( & \widetilde{\langle S^z \rangle}(\delta k' - \delta k) \;
    c^\dagger_{\delta k,r} c_{\delta k',r}  
    + 
    \widetilde{\langle S^z \rangle}(\delta k' - \delta k) \;
    c^\dagger_{\delta k,l} c_{\delta k',l} \nonumber \\
    & + 
    \widetilde{\langle S^z \rangle}(\pi+(\delta k' - \delta k)) \;
    c^\dagger_{\delta k,r} c_{\delta k',l} 
    + 
    \widetilde{\langle S^z \rangle}(\pi+(\delta k' - \delta k)) \;
    c^\dagger_{\delta k,l} c_{\delta k',r} \Bigg)
    - \sum_i \frac{\langle S^z_i \rangle}{2}
    \label{eq:pert_term}
\end{align}
\end{widetext}
where
$\widetilde{\langle S^z \rangle}(p) \equiv
\sum_i e^{i p r_i} \langle S^z_i \rangle$ is the (discrete)
Fourier transform of the off-chain spin configuration.
$r_i$ is understood to be a set of integers say from
$0$ to $L-1$, etc. Also
$\widetilde{\langle S^z \rangle}(p)
= \widetilde{\langle S^z \rangle}(-p)$ since 
$\langle S^z \rangle \in \Re$.
Now we ask for which conserved off-chain spin configuration
do we gain the most energy upon including the perturbation.
The first two terms in Eq.~\ref{eq:pert_term} connect modes
with same helicities, whereas the last two terms
in Eq.~\ref{eq:pert_term} connect modes with different helicities.
It is intuitively clear that the most energy to be gained
is when the perturbation term from the off-chain configuration
connects modes that are closest in energy (ideally
degenerate energy modes either with same or 
different helicity). 
The best that we can do from this point of view is to
either have a ferromagnetic off-chain configuration for
which $\widetilde{\langle S^z \rangle}(p=0) \neq 0
\; (= \frac{L}{2})$ and
all other $\widetilde{\langle S^z \rangle}(p) = 0$, or 
an antiferromagnetic off-chain configuration for which
$\widetilde{\langle S^z \rangle}(p=\pi) \neq 0
\; (= \frac{L}{2})$ and
all other $\widetilde{\langle S^z \rangle}(p) = 0$.
If we take the ferromagnetic case with
$\langle S^z_i \rangle = \frac{1}{2}$, then the
net many-body energy gain due to shifts in
the one-particle energies is 
$J_{\text{Z-I}} \left(\sum_{\epsilon(p)+\frac{J_{\text{Z-I}}}{2}<0} \frac{1}{2}-\sum_{-\Lambda < \delta k < \Lambda} \sum_{\alpha=\{l,r\}} \frac{1}{4}
\right)$ which is lower bounded 
by $-J_{\text{Z-I}}\sum_{-\Lambda < \delta k < \Lambda} \sum_{\alpha=\{l,r\}} \frac{1}{4}
= -\sum_{-\Lambda < \delta k < \Lambda} \frac{J_{\text{Z-I}}}{2}$.
For this we keep in mind that the counting of modes should
remain the same, i.e. $\sum_{\delta k} \sum_{\alpha=\{l,r\}} = \sum_i$
which leads to $\sum_i \frac{\langle S^z_i \rangle}{2}
= \sum_{\delta k} \sum_{\alpha=\{l,r\}} \frac{1}{4}$.

If we take the antiferromagnetic case, the 
last real-space sum term in Eq.~\ref{eq:pert_term} is 
now equal to $\sum_i \langle S^z_i \rangle=0$. 
The effect of the perturbation is now to open a gap
at the Fermi points in the single-particle spectrum
which ``pushes down" all energy eigenvalues below
the Fermi energy (by
different amounts), and ``pushes up" all energy eigenvalues
above the Fermi energy. This leads to a decrease in
the many-body ground state energy which is again the 
half-filled spinless Fermi sea but now with a gap to
(single-particle and multi-particle) excitations. The
many-body energy gain will generically be 
of the form $-\sum_{\Lambda <\delta k <\Lambda} 
\sqrt{\frac{J_{\text{Z-I}}}{2}^2+(\ldots)^2}$
where $\ldots$ represent the energy difference of the 
close-in-energy modes of different helicities separated by a 
momentum of $\pi$ which are now hybridized by the perturbation
term in Eq.~\ref{eq:pert_term}. This is a greater many-body energy 
gain than the ferromagnetic off-chain configuration.
This is how the antiferromagnetic off-chain
spin configuration gets selected in the XY limit. 
The above can be made rigorous by making
rigorous the intuitive step taken above of ``connecting
modes closest in energy through the perturbation to gain
the most reduction in the (unperturbed) many-body ground 
state energy". This is
an open and a mathematical physics 
question~\cite{Chakraborty_Pujari} even though the physical
reason for the ground state sector selection is clear.
\begin{figure*}
\centering
\includegraphics[clip=true,trim=0 0 24cm 0, width=0.3\linewidth]{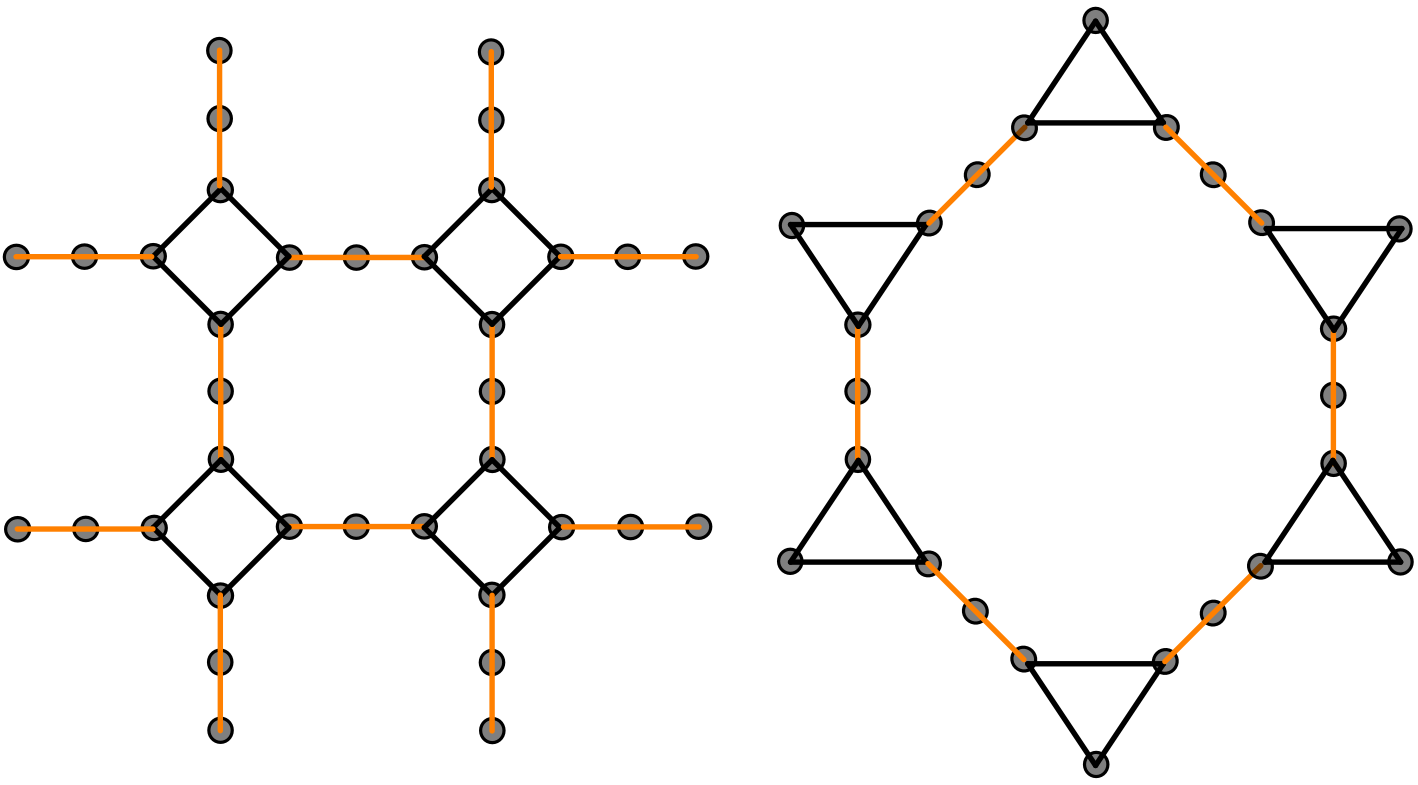}
\;\;\;\;\;\;\;\;\;\;\;\;
\includegraphics[clip=true,trim=26cm 0 0 0,width=0.3\linewidth]{non_stripy_eg}
\caption{\label{fig:non_stripy} Some
additional constructions not based on the
one-dimensional chain geometry.}
\end{figure*}

For the non-bipartite XY case, the lattice Hamiltonian is
\begin{equation}
    H = J_{\text{XY}} \sum_{\langle i,j \rangle} c^\dagger_i c_j + \text{h.c.}
        + J_{\text{Z-I}} \sum_i \left( n_i + n_{i+1} - 1 \right) \langle S^z_i \rangle  
\end{equation}
Upon redoing the steps above, we get
\begin{widetext}
\begin{align}
    \sum_i \langle S^z_i \rangle & \left(n_i + n_{i+1}- 1\right) = 
    \sum_{\delta k,\delta k' < \Lambda} 
    \Bigg( \left( 1 + e^{i (\delta k' - \delta k)} \right) \widetilde{\langle S^z \rangle}(\delta k' - \delta k) 
    \;
    c^\dagger_{\delta k,r} c_{\delta k',r}  
    + 
    \left( 1 + e^{i (\delta k' - \delta k)} \right) \widetilde{\langle S^z \rangle}(\delta k' - \delta k) \;
    c^\dagger_{\delta k,l} c_{\delta k',l} \nonumber \\
    & + 
    \left( 1 - e^{i (\delta k' - \delta k)} \right) \widetilde{\langle S^z \rangle}(\pi+(\delta k' - \delta k)) \;
    c^\dagger_{\delta k,r} c_{\delta k',l} 
    + 
    \left( 1 - e^{i (\delta k' - \delta k)} \right)\widetilde{\langle S^z \rangle}(\pi+(\delta k' - \delta k)) \;
    c^\dagger_{\delta k,l} c_{\delta k',r} \Bigg)
    - \sum_i \langle S^z_i \rangle
    \label{eq:pert_term_kagome}
\end{align}
\end{widetext}
By using the earlier intuition, now the helicity cross-connecting
terms due to the perturbation actually cancel out due to
the $\left( 1 - e^{i (\delta k' - \delta k)} \right)$ term in
the antiferromagnetic off-chain spin configuration sector. Thus the
ferromagnetic off-chain spin configuration is selected as the
ground state sector perturbatively in the XY limit in presence
of triangular motifs.

\subsection{Bipartite X-Ising}
\label{subsec:unfrus_XX}

This last case is the most unusual and interesting.
It is the X-Ising variant on lattices without triangular motifs. The ground state here turns out to be an Ising ordered/disordered state for the on-chain spins coexisting with a very short-ranged liquid-like (classical) paramagnetic state for the off-chain spins! In other words, all the conserved off-chain spin sectors have the same spectrum as indicated in Fig.~\ref{fig:unfrustrated_xy_xx}. This arises due to an additional structure present in this case that was absent in the previous cases. The structure is the presence of additional extensively large conserved quantities apart from the off-chain spins that are mutually anticommuting when there is a shared site. General consequences of this structure form the subject of Ref.~\cite{Ising_theorem_preprint}. The additional conserved quantities are products of three X-component operators from sites on each of the Z-Ising segments (orange solid lines in Fig.~\ref{fig:unfrustrated_xy_xx}), \unhighlight{i.e. $S^x_i S^x_j S^x_k$ where $\{i,j,k\}$ forms a Z-Ising segment $J_{\text{Z-I}} \left(S^z_i S^z_j + S^z_j S^z_k\right)$ that is \emph{disconnected} from other Z-Ising segments.} Since there are an extensive number of these operators coming from all the Ising segments, this leads to the extensive degeneracy associated with the off-chain spin configurations. The extensive scaling with system size of these additional conserved quantities can depend on the model, i.e. here they scale as $L^2$ same as the conserved off-chain spins. There can be variants where they scale as $L$ for example~\cite{Nussinov_Ortiz_2023}. This should not be thought of as violating the third law of thermodynamics~\cite{third_law}, but rather is well-understood as capturing the relevant physics of real systems operating at the relevant temperature scales in a physical setting or experiment~\cite{Masanes_Oppenheim_2017} similar to classical spin ices~\cite{spin_ice_review}. In presence of other (even smaller) couplings, a natural guess is to presume the existence of a (possibly gapless) “slow” mode primarily composed of the off-chain spins as excitations on top of the ground state, apart from the gapped Majorana excitations from the on-chain spins which will be higher in energy compared to the presumed slow mode. The solvable limit can be said to have an extensive number of zero modes from the perspective of this expectation. The above structure also implies that the absence of triangular motifs is not an absolute requirement for this physics; rather one only requires Z-Ising (closed or open) segments with at least one conserved spin in the spirit of the above constructions connected through solely X-Ising (or solely YY-Ising) segments as shown in Fig.~\ref{fig:non_stripy}. \unhighlight{Also must be mentioned before proceeding further, that the results discussed above pertaining to the particular chain-based construction of Fig.~\ref{fig:unfrustrated_xy_xx} is essentially a rediscovery and restatement of the result found in Ref.~\cite{Nussinov_Ortiz_2008}. The authors of Ref.~\cite{Nussinov_Ortiz_2008} arrived at them with different motivations coming from orbital physics in transition metal systems and did not discuss explicitly the classical spin liquid aspect of the off-chain spins as well as the phenomenological consequence analogous to the Kitaev honeycomb model that we describe next.}

\begin{figure*}
    \centering
    \includegraphics[width=0.72\linewidth]{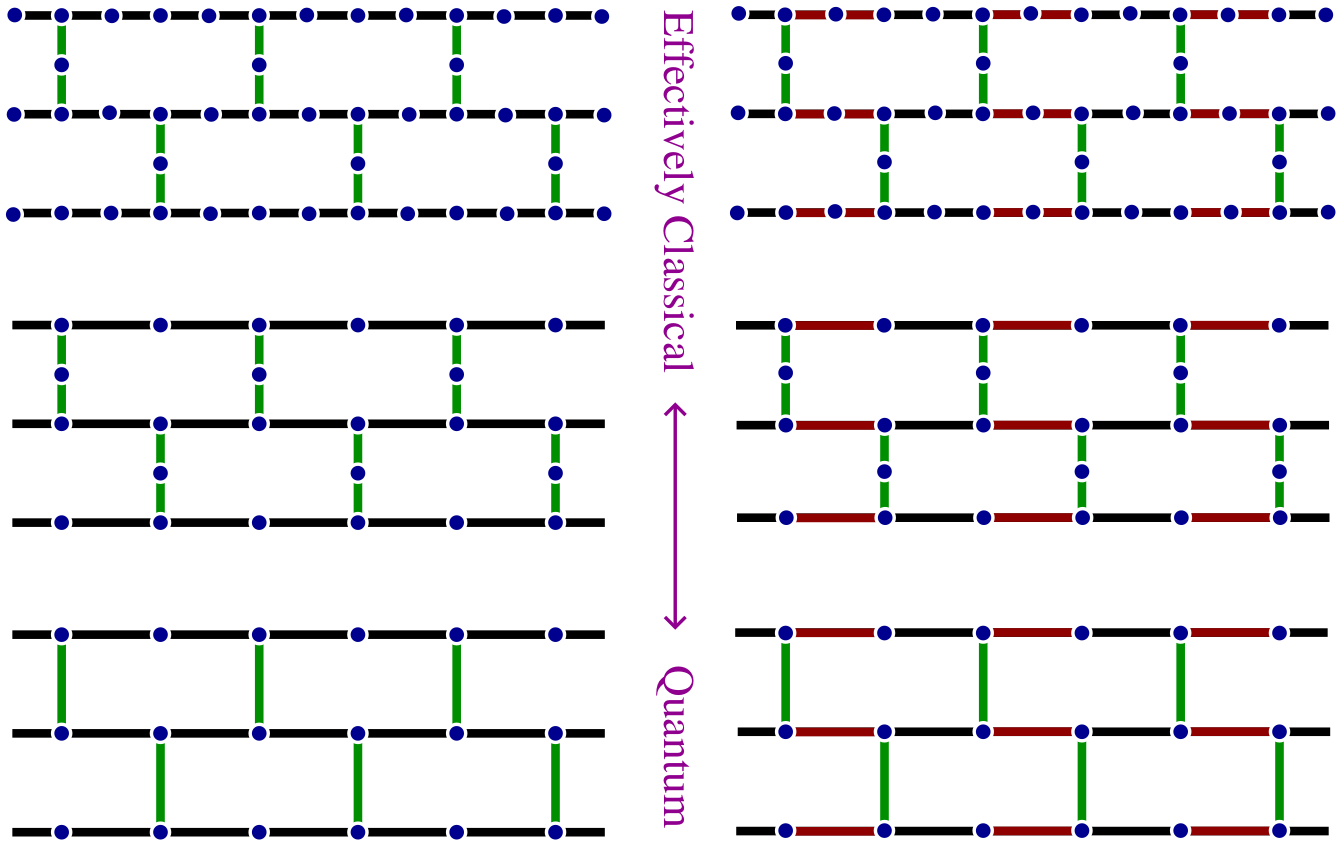}
    \caption{A top-to-bottom sequence of models with
    increasing levels of quantum behavior. 
    \unhighlight{Black lines represent X-Ising couplings. Red lines represent
    YY-Ising couplings. Green lines represent Z-Ising
    couplings. Blue dots represent $S=\frac{1}{2}$ degree of freedom.}
    Left sequence has the model of 
    Fig.~\ref{fig:unfrustrated_xy_xx} (b) at the middle.
    Right sequence has the Kitaev
    model at the bottom.  The top effectively classical model 
    in this sequence was studied by 
    Mizoguchi~\cite{Mizoguchi_2022}. Lieb lattice
    variants of the top row of brickwork lattices
    can also be envisaged.}
    \label{fig:classical_to_Kitaev}
\end{figure*}

Given the above massively degenerate ground state manifold,  
one now expects a two-step entropy release
similar to that in the Kitaev spin liquids 
in presence of thermal fluctuations
~\cite{Nasu_Udagawa_Motome_2014,Nasu_Udagawa_Motome_2015}
with residual entropy effects as temperature goes to zero.
The first step would correspond to the
entropy release due to the off-chain spins 
and the second step to the on-chain spins. The
relative magnitude of the two steps will depend on the ratio 
of the number of off-chain spins to on-chain
spins with it being 1:2 for the lattice geometry
in this case (Fig.~\ref{fig:unfrustrated_xy_xx}). 
Thus these states can be considered as simpler
variants of the Kitaev spin liquid.
As differences, the conserved off-chain
spins are geometrically independent and site-local
in our constructions, whereas
in the Kitaev model, the conserved plaquette operators are
``interlocked" with common sites. The excitations also have
a stripy nature in all of the above 
models, whereas the 
Majorana excitations in the Kitaev spin liquid are fully
two-dimensional in nature. 
One can have non-stripy constructions as well
with localized excitations throughout as shown
Fig.~\ref{fig:non_stripy}.
\unhighlight{The conserved quantities of the Kitaev model manifest
as $Z_2$ flux background for these Majorana fermions. Being 
fluxes in two dimensions, they are relevant for the energetics
of the excitations. In the model considered here, the conserved 
off-chain
spins only lead to gauge artifacts for the stripy Majorana
excitations. Thus the gauge artifact argument described 
in the beginning gives another perspective 
on the extensive degeneracy of this model.}
If there are additional
conserved sites on the on-chain X-Ising spin chains,
then the resultant models will be rendered effectively
classical. 
The models considered here can therefore be said to be
``in between" effectively classical states
and fully quantum spin liquids 
as shown in Fig.~\ref{fig:classical_to_Kitaev}.

\section{Summary and Discussion}
\label{sec:conclu}

\unhighlight{
In this paper, we constructed a host of frustrated bond-dependent
$S=\frac{1}{2}$ models in two (and higher) dimensions 
that are solvable by the Jordan-Wigner 
transformation~\ref{sec:model}.
We focused on the ground state properties
and spectral computations for the constructed models
in Sec.~\ref{sec:results}.
The various cases were categorized as:
1) Bipartite XY constructions as in Sec.~\ref{subsec:unfrus_xy}
which hosts a gapped Ising antiferromagnetic ground
state with gapped on-chain spinon
excitations. 2) Non-bipartite XY and X-Ising constructions as in
Sec.~\ref{subsec:frus_xy_xx} which host ferromagnetic
states along with, respectively, on-chain Luttinger liquid
states with gapless spinon excitations, and on-chain Ising ordered or disordered
states with gapped Majorana excitations away from the quantum critical 
point $\left( \frac{J_{\text{Z-I}}}{J_{\text{X-I}}}=0.5 \right)$. 3) Bipartite X-Ising constructions as in 
Sec.~\ref{subsec:unfrus_XX} which hosts an unusual
coexistence state of a classical spin liquid (with extensive number of 
``zero modes'') on the
off-chain sites and quantum Ising order or disorder
on the on-chain sites. Details on the numerical
methods and perturbative arguments for ground state
sector selection within the solvable parameter regime
were given in Secs.~\ref{subsec:numerics} and 
~\ref{subsec:pert_arg}.
}

One can also calculate on-chain transverse 
spin-spin correlations $\langle \sigma^+_i \sigma^-_j
+ \text{ h.c.}\rangle$
using known techniques~\cite{Tonegawa_SSC_1981}.
Computation of the full partition function seems
non-trivial, especially summing over the conserved
off-chain spin sectors and is an open question.
The solvability is also preserved in presence
of magnetic field along appropriately chosen
quantization axes.
It is desirable to study various other aspects 
of these models such as 1) thermodynamical
quantities at finite temperatures, 2) 
dynamical phenomena in presence of quenches and drives,
and 3) investigation of the stability of the reported
magnetic states outside the solvable parameter
regime. This is left for the future. We comment on this 
last issue some more in what follows.

\begin{figure*}
    \centering
    \includegraphics[clip=true,trim=0 0 17cm 0, width=0.3\linewidth]{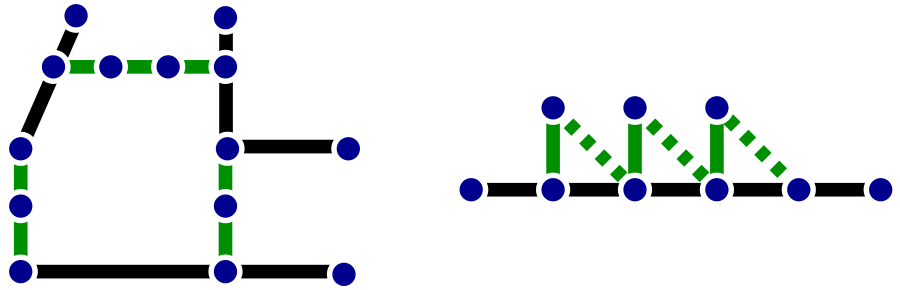}
    \;\;\;\;\;\;\;\;\;\;\;\;
    \includegraphics[clip=true,trim=16cm 0 0 0, width=0.3\linewidth]{bus_architecture.png}
    \caption{Some imagined quantum architectures 
    \unhighlight{following the
    color scheme of Fig.~\ref{fig:classical_to_Kitaev}.
    The dashed green Z-Ising couplings can be switched on or off.}
    }
    \label{fig:bus_architecture}
\end{figure*}

\subsection{Stability considerations}
\label{subsec:stability}

It is also important to consider the stability
of the grounds states discussed here with respect to
natural deformations of the Hamiltonian, i.e. off-chain
XY couplings and on-chain Z-Ising couplings. For the
XY variants with on-chain Z-Ising couplings as deformations, 
one can invoke Luttinger liquid arguments to argue for 
stability~\cite{Giamarchi_book,Sachdev_book}.
The bipartite XY case would be stabler to on-chain
Z-Ising deformations due to the presence of the gap in the
on-chain spinless fermionic excitations. The same would
be expected for the non-bipartite and bipartite X-Ising case
due to the gap in the on-chain spectrum in presence of
Ising order or disorder. 
Stability to off-chain deformations in the first three
cases of bipartite XY, non-bipartite XY and X-Ising models
is evidenced by the presence of the sector gap as shown in
Figs.~\ref{fig:unfrustrated_sector_gaps},~\ref{fig:frustrated_sector_gaps}. This is because it is precisely these 
deformations such as off-chain XY or X-Ising couplings 
that would lead to quantum fluctuations in the off-chain spin 
Z-values that were otherwise conserved by construction.
\unhighlight{Presence of the sector gap suggests that if the
energy scale associated with these fluctuations is lesser
than the gap, then the ground states should be stable
till they become large enough. 

For the particular case of non-bipartite XY construction with 
gapless on-chain excitations, one may worry that the inter-chain
interactions induced through the gapped off-chain excitations
--- by integrating out the off-chain spin degrees of
freedom --- may destabilize the system and lead to a gapped
state for the on-chain spectrum as well~\cite{Schnyder_Starykh_Balents_2008,Starykh_Balents_2007}. 
However, given 
the perturbative stability arguments in both 
$\frac{J_{\text{XY}}}{J_{\text{Z-I}}} \ll 1$ and 
$\frac{J_{\text{XY}}}{J_{\text{Z-I}}} \gg 1$ limits for the gapless
on-chain state within the solvable bond-dependent
parameter regime as discussed previously in Sec.~\ref{subsec:pert_arg}, 
we conjecture that additional off-chain perturbations will lead
to a locking of the spin superfluid order parameter across
the chains in the ground state. This suggests that
 in presence of off-chain perturbations a long-range
ordered spin superfluid ground state -- without a gap 
in the thermodynamic limit -- obtains in the
non-bipartite XY construction up to a finite range of such 
local perturbations, outside of which a gap may open. 
Even in the solvable regime, one may wonder if finite 
(low) temperatures can lead to a locking of the spin 
superfluid order across the chains via an 
order-by-disorder mechanism.}

\unhighlight{Note that the excitations on top of a 
long-ranged ordered spin superfluid ground 
state in two dimensions are bosonic magnons.
Even so, the fermionic spinon excitation physics discussed 
in this work can play a role in a ``dimensional reduction'' sense,
i.e. in an appropriate regime of temperature, the coupling 
ratio $\frac{J_{\text{XY}}}{J_{\text{Z-I}}}$ and other couplings
(including an external magnetic field in Z direction)
that take us out of the solvable regime, the effective description
may still be in terms of on-chain
spinon excitations and localized spin flip excitations
with majority weight on the off-chain spins. A recent 
experimental example of a quasi-two-dimensional 
copper based quantum magnet
showing such dimensional reduction phenomenology can
be found in Ref.~\cite{Reinold_etal_2025}.
See also Ref.~\cite{Shao_etal_2017} in the context of 
two-dimensional $SU(2)$-symmetric Heisenberg magnets
with magnonic excitations where ``where a broad 
spectral-weight continuum at wave vector $q=(\pi,0)$ 
was interpreted as deconfined spinons'' in relation
to neutron-scattering experiments on another copper
based compound~\cite{Dalla_Piazza_etal_2014}.}

Explicit verification of the above arguments may
be done using various Quantum Monte Carlo or Matrix product
based methods.
In the case of the bipartite X-Ising model with an
extensive degeneracy from the paramagnetic off-chain spins, 
the off-chain deformations would be a possible mechanism 
for the development of the conjectured slow mode
and the associated restoration of the third law 
of thermodynamics.

\subsection{Physical realization} 
\label{subsec:physical_realization}
We end with some remarks
with a view towards physical
realization. \unhighlight{Firstly, the models discussed here
in the JW-mapped fermionic form are reminiscent
of Falicov-Kimball models where also itinerant 
electrons or Bloch waves interact with 
static or localized electrons. They 
have been invoked to study a variety of phenomena
such as mixed-valence and metal-insulator transitions.
More details on these electronic phenomenan can
be found in the following reviews~\cite{Falicov_Kimball_review_1994,Falicov_Kimball_review_1998,Falicov_Kimball_review_2001,Falicov_Kimball_review_2003}. 
See this Ref.~\cite{Li_Chen_Ng_2019} for an example
of Falicov-Kimball models with Majorana excitations.
The models discussed in this paper realize
aspects of Falicov-Kimball physics, however and it must be
emphasized, in a non-electronic context and only after
a \unhighlight{\emph{non-local}} transformation. 
In other words, the
spin-spin correlations encode Falicov-Kimball type
effects involving the itinerant electrons in a highly 
non-linear way. Furthermore, the possible geometries
allowed by the embedding mechanism laid down here 
admits on-site potentials that act as a ``shared'' 
background for sites on different chains that are somewhat 
unnatural in the
electronic context that motivated the Falicov-Kimball
model in the first place, i.e. ``as a set of localized
states centered at the sites of the metallic ions
in the crystal''~\cite{Falicov_Kimball_1969}.}

In a fully localized context without any
itinerant electrons, 
one may imagine physical realizations such as an 
insulator material with
localized $S=\frac{1}{2}$ moments with a crystal structure
and quantum chemistry that accommodates a grid structure of
easy-axis and easy-plane spin chains as the case may be.
If the non-bipartite XY case could be so realized, it would
allow for highly anisotropic spin transport along the
spin chains which can have practical applications such
as in spin-based classical technologies.
This could be an interesting avenue of
exploration for the materials design and search community
as alternative to the Kitaev material class with
bond-dependent magnetic couplings.

But perhaps the more natural application of the ideas
laid out in this work may be in the realm of quantum
technologies involving artificial platforms with controllable
qubits. The grid structure proposed here can provide
a blueprint for a bus architecture for quantum information
traffic. The conserved off-chain spins can sequester the
spin information to the on-chain spin chains.  
Some imagined architectures are shown in 
Fig.~\ref{fig:bus_architecture}. For example the right
panel of Fig.~\ref{fig:bus_architecture} could be an
use case where turning on and off the dashed Z-Ising couplings
would act as a valve for spin information flow.
Interestingly, such an architecture has been recently
reported in Ref.~\cite{Kim_Kandala_etal_2023} with a
different application as its focus.

\vspace{2mm}

\prlsec{Acknowledgements}
I thank Sumilan Banerjee, Ganapathy Baskaran, Soumya Bera, Subhro Bhattacharjee, Hitesh Changlani, Kedar Damle, Nisheeta Desai, Giniyat Khaliullin, Gregoire Misguich, Laura Messio, Subroto Mukherjee, Ganpathy Murthy,  Hridis Pal and Sumathi Rao for discussions. I thank Subhro in particular for the perturbative argument in the $J_{\text{Z-I}}/J_{\text{XY}} \ll 1$ limit. I am grateful to the Toulouse strongly correlated electron systems group, Laboratory of Theoretical Physics, University Paul Sabatier for the hospitality and discussions during a part of the work. I especially thank Fabien Alet for encouragement and feedback. Funding support from SERB-DST, India via Grant No. MTR/2022/000386 and partially by Grant No. CRG/2021/003024 is acknowledged. I acknowledge the hospitality and support of ICTP, Trieste during a part of the work. I thank CEPIFRA during a part of this work for support to the ``Indo-French workshop on Topology and Entanglement in Quantum Matter 2024'' held at the University of Toulouse, France between June 3rd-7th, 2024.

\bibliography{JWrefs}


\end{document}